\def\references{\subsection*{REFERENCES}
\bgroup\parindent=0pt\parskip=\itemsep
\def\refpar{\par\hangindent=1.2em\hangafter=1}}
\def\endreferences{\refpar\egroup}
\def\refpar{\relax}
\def\reference{\relax\refpar}
\def\plotfiddle#1#2#3#4#5#6#7{\centering \leavevmode
\vbox to#2{\rule{0pt}{#2}}
\includegraphics{#1}}

\def\lunits{$\rm erg~s^{-1}~$}
\def\funits{$\rm erg~cm^{-2}~s^{-1}~$}
\def\xrbunits{$\rm keV~cm^{-2}~s^{-1}~sr^{-1}~keV^{-1}$}
\def\ns{{$\log N-\log S$ }}
\def\asca{{\it ASCA}~}
\def\ginga{{\it Ginga}~}

\def\simless{\buildrel < \over {_\sim}} 
\def\simgreat{\buildrel > \over {_\sim}} 

\documentstyle{mn}

\begin{document}

\title[{\it ASCA} observations of deep {\it ROSAT} fields I]
     {{\it ASCA} observations of deep {\it ROSAT} fields I. the nature of
     the X-ray source populations}

\author[I. Georgantopoulos et al.]
     {I. Georgantopoulos$^1$, G.C. Stewart$^1$, A.J. Blair$^1$
     T. Shanks$^2$, R.E. Griffiths$^3$,
\newauthor
     B.J. Boyle$^4$, O. Almaini$^5$, N. Roche$^6$  \\
$^1$ Department of Physics \& Astronomy, The University of Leicester,
     Leicester LE1 7RH \\
$^2$ Physics Department, University of Durham, South Road,
     Durham DH1 3LE \\
$^3$ Department of Physics, Carnegie Mellon University, Wean Hall,
     5000 Forbes Ave., Pittsburgh, PA 15213 U.S.A. \\
$^4$ Anglo-Australian Observatory, PO Box 296, Epping NSW 2121, Australia \\
$^5$ Institute of Astronomy, Madingley Road, Cambridge CB3 0HA \\
$^6$ Johns Hopkins University, Department of Physics \& Astronomy,
     Baltimore, MD 21218, U.S.A.
}

\maketitle

\begin{abstract}

We present \asca GIS observations (total exposure $\sim$100-200 ksec)
of three fields which form part of our
deep {\it ROSAT} survey. We detect 26 sources down to a limiting flux (2-10
keV) of
$\sim5\times10^{-14}$ \funits. Sources down to this flux level
contribute $\sim$30 per cent of the 2-10 keV
X-ray background. The number-count distribution, \ns, is a factor
of three above the {\it ROSAT} counts, assuming a spectral index of
$\Gamma=2$ for the {\it ROSAT} sources.  This suggests the presence at
hard energies
of a population other than the broad-line AGN which contribute to
the {\it ROSAT } counts. This is  supported by spectroscopic observations
that show a large fraction of sources that are  not obvious
 broad-line AGN. The average 1-10 keV spectral index of these sources
is flat  $\Gamma = 0.92\pm0.16$, significantly different than
that of the broad-line AGN ($\Gamma= 1.78\pm0.16$).
Although some of the Narrow Emission Line Galaxies which are detected
with {\it ROSAT} are also detected here, the nature of the flat
spectrum sources remains as yet unclear.
\end{abstract}

\vspace{1.5cm}

\section{INTRODUCTION}

The origin of the diffuse X-ray emission, the X-ray background (XRB),
that dominates the sky from energies of 0.1 keV up to 1 MeV remains uncertain.
The bulk of the XRB  cannot originate from hot intergalactic gas
(Mather et al. 1990)  but instead must arise in
discrete sources (for a review see  Fabian \& Barcons 1992).
At soft energies ($<$2 keV) great strides have been made after the launch
of the X-ray satellite {\it ROSAT} (Tr\"umper 1990). Deep observations
with the {\it ROSAT} PSPC (Shanks et al. 1991, Hasinger et al. 1993,
Branduardi-Raymont et al. 1994, Georgantopoulos et al. 1996)
reveal a high density of X-ray sources ($>$400 $\rm deg^{-2}$,
at $2\times 10^{-15}$ \funits),
which contribute over than half of the  soft (0.5-2 keV) XRB.
The integral number-count distribution, logN-logS,
turns over to a flatter  than Euclidean power law slope
 at $S_{0.5-2 keV} \approx 2\times 10^{-14}$ \funits, tending to
 a slope of $\gamma \sim 1$  (Hasinger et al. 1993, Vikhlinin et al. 1995).
 Spectroscopic follow-up observations have shown that the majority of
the  sources are broad-line, type I AGN, ie QSOs and Seyfert 1 galaxies, at a
mean
 redshift of z=1.5 (e.g. Shanks et al. 1991, Boyle et al. 1995,
Carballo et al. 1995, Georgantopoulos et al. 1996, Bower et al. 1996).
However, the QSO luminosity function, the anisotropy of the
XRB and the average QSO spectra argue strongly against a QSO
origin for  the soft XRB.
The QSO luminosity function and its evolution has been  derived using
combined {\it Einstein} and {\it ROSAT} data (Boyle et al. 1993, 1994).
An integrated QSO contribution of only $\sim$ 50 per cent in the 0.5-2 keV band
is
determined.
A similar conclusion is reached from studies of the XRB anisotropy.
The auto-correlation function (ACF) of the 1-2 keV XRB presents a weak
signal (Georgantopoulos et al. 1993, Soltan \& Hasinger 1994, Chen et al.
1994) which lies below the strong ACF signal predicted from the
optical  QSO correlation function  (Shanks \& Boyle 1994,
Georgantopoulos \& Shanks 1994).
Finally,  the average QSO spectra in  deep {\it ROSAT} fields
 have a photon spectral index of $\Gamma \sim 2$ (Stewart et al.
1994, Almaini et al. 1996), steeper than the spectrum of the XRB
($\Gamma \sim 1.5$) at soft energies (Georgantopoulos et al. 1996).
This extends the spectral paradox already noted in harder X-rays (Boldt
1987) and suggests either
a population with  a flat spectral index
or one which is heavily absorbed and remains unidentified  at faint fluxes.
Indeed, {\it ROSAT} PSPC exposures reveal  a new population of X-ray
luminous ($\rm L_x \simgreat 10^{42}$ \lunits) optically faint galaxies,
(Roche et al. 1995, Griffiths et al. 1995, Boyle et al. 1995, Carballo et al.
1995,
Georgantopoulos et al. 1996, Griffiths et al. 1996)
which do not have the broad emission lines typical of QSOs.
Although these narrow emission line galaxies (NELGs) are too faint for
individual
X-ray spectral analysis, their co-added spectra appear to be flat
($\Gamma \sim 1.5$), similar to the XRB spectrum in the same
energy band (Almaini et al. 1996, Romero-Colmenero et al. 1996).
Strong positive cross-correlation signals between the PSPC background
fluctuations and faint galaxies (B$<$23) have shown that these
contribute a significant fraction (at least 17 per cent) of the
soft XRB (Roche et al. 1995).

On the other hand, the hard XRB ($>$ 2 keV), where the bulk energy density
resides, remains less well explored,
as measurements at hard X-rays have been performed mainly  using
collimated X-ray detectors with  coarse (degrees) angular resolution.
The {\it HEAO-1} experiment (Wood et al. 1984) has detected several hundred
sources
over the whole sky, in the 2-10 keV band,
with  fluxes $\simgreat 5\times 10^{-12}$ \funits, contributing less than
5 per cent  of the hard XRB intensity.
The logN-logS from {\it HEAO-1} (Piccinotti et al. 1982) and {\it Ginga} (Kondo
1990)
is represented by a Euclidean power law with a normalization a factor of
2-3 above that  of the {\it ROSAT} logN-logS.
The fluctuations analysis of the
hard XRB in {\it Ginga} fields (Butcher et al. 1997)
extends these conclusions down to flux levels of $5\times 10^{-13}$
\funits.
 These imply that a
flat spectrum ($\Gamma \simless 1.5$) or absorbed  population
($N_H>3\times 10^{21}$ $\rm cm^{-2}$) dominates the hard energies
 (e.g. Ceballos \& Barcons 1996).
 The majority of the bright hard X-ray sources are
nearby type I AGN (Piccinoti et al. 1982).
They have  a power law spectrum of $\Gamma \sim 1.7$
(e.g. Nandra \& Pounds 1994) inconsistent
with  the XRB spectrum  in this band which has a spectral index of
$\Gamma \sim 1.4$ (Marshall et al. 1980, Gendreau et al. 1995).

The launch of the X-ray satellite
{\it ASCA} provides  the first opportunity to observe  the hard (2-10 keV)
X-ray sources down to a flux level of few times $10^{-14}$ \funits,
 about two orders of magnitude fainter than the {\it HEAO-1} survey, but still
 an order of magnitude above the flux limit of the deepest {\it ROSAT} surveys.
Here, we present the results from \asca observations of three
fields included in our deep {\it ROSAT} survey (Shanks et al. in preparation).
The benefits of observing {\it ROSAT} fields with previous spectroscopic
follow-up observations  are obvious, as we can immediately obtain the optical
identifications for many of the \asca sources.
The major aim of this paper is to examine the nature of the
the faint hard X-ray sources and to estimate their contribution to the
XRB. First, we discuss the X-ray and optical properties of the
detected sources and then we derive  their number-count distribution,
logN-logS, as well as their contribution to the hard XRB.

\section{THE X-RAY OBSERVATIONS}

\subsection{Data reduction}

Our deep {\it ROSAT}  survey (Shanks et al.; in preparation)
consists of 7 PSPC fields with exposure times up to 80 ksec and covers $\sim$
2 $\rm deg^{2}$. About 300 sources have been detected down to a flux limit
of $3\times 10^{-15}$ \funits (0.5-2 keV) in the central 20 arcmin radius
of the PSPC field-of-view where the detector/telescope sensitivity is the
highest.  Both the optical and
the X-ray observations from the first 5 fields are described in detail in
Georgantopoulos et al. (1996).

Three fields from our {\it ROSAT} survey (QSF3, GSGP4, BJS855) have been
observed with the \asca satellite (Tanaka et al. 1994).
\asca was launched in February 1993
and carries two SIS (Solid State Imaging Spectrometer)
and two GIS (Gas Imaging Spectrometer) each with its own X-ray telescope
(XRT) (Serlemitsos et al. 1995). The SIS instruments cover a field-of-view
of approximately 20x20 arcmin whilst the GIS instruments cover an area
of 20 arcmin radius. Here we present the analysis of the GIS data alone
 because a) the GIS field-of-view matches that
used in our {\it ROSAT} survey and b) with the GIS we maximize the
effective
exposure times, ie the net exposure times after rejecting time periods
with high rates of particle events.
In table 1 (please note that this table will not be distributed before
the actual publication of the paper) we give the field names in column (1);
 equatorial coordinates (J2000),
 in columns (2) and (3); the hydrogen column density in units of
 $10^{20}$ $\rm cm^{-2}$ (Stark et al. 1992) in column (4);  finally  the
effective exposure times per telescope in ksec are given in column (5).
 The QSF3 field was observed four times during the period of
Performance Verification (PV phase). The first observation was in June
1993 and the remaining three in September 1993.
The GSGP4 and the BJS855 fields were observed in June 1994 and November
1995 respectively.

Images are created in sky coordinates using the FTOOLS/XSELECT software
(Day et al. 1995). We reject a small fraction of the data
that corresponds to times of
high particle background, keeping only data which satisfied the
following selection criteria: a) elevation angle from the Earth limb
greater than 5 degrees b) the satellite remains outside the South
Atlantic Anomaly c) the Radiation Belt Monitor gives values below
200 $\rm ct~s^{-1}$. Finally, a bright ring around the edge
of the field-of-view that contains mostly particle background events
is removed from the image (see Day et al. 1995).

The nominal energy response of the GIS+XRT combination is  0.8-12 keV.
However, below 1 keV and above 10 keV the response drops  rapidly.
 Here, we use the 2-10 keV band for our  source detection.
The 1-2 keV overlaps with the {\it ROSAT} PSPC energy response and
it is used to check the {\it ASCA} results against the well calibrated
{\it ROSAT} data.
The Point Spread Function (PSF) of the GIS+XRT combination has a
half-power-radius of 1.5 arcmin on-axis. The radius of the encircled energy
fraction depends  on the off-axis angle. The 2 arcmin radius
includes $\sim$60 per cent of the source light on-axis while at 17 arcmin
this fraction reduces to $\sim40$ per cent (e.g. Takahashi et al. 1995).

\begin{table}
\begin{center}
\caption{ List of ASCA fields}
\vskip 0.5cm
\begin{tabular}{ccccc}
\small
Field & $\alpha$ & $\delta$ & $N_H $ & Exposure  \\
      &          &          &  $(10^{20}\rm cm^{-2})$ & (ksec) \\   \hline
QSF3  & 03 41 44.4 & -44 07 04.8  & 1.7 & 109 \\
GSGP4 & 00 57 25.2 & -27 37 48.0  & 1.8 & 50  \\
BJS855&10 46 24.0  & -00 20 38.4  & 1.8 & 54  \\
\end{tabular}
\end{center}
\end{table}

We mosaic the images from the two detectors GIS2 and GIS3 in order to
increase the exposure time and hence to maximize the signal-to-noise
ratio for source detection.
As the optical axes of the two telescopes do not coincide,
 the maximum exposure times are not simply double the exposure times
 given in table 1.  The maximum exposure times in the mosaic
fields are approximately 185, 95 and 100 ksec for
QSF3, GSGP4 and BJS855 respectively.
The background appears to be  uniform,  within
17-18 arcmin radius, despite the strong vignetting of the
XRT telescope (only $\sim$ 30 per cent of the
light is captured  at an off-axis angle of 18 arcmin,
Serlemitsos et al. 1995).
The  lack of vignetting in the images is attributed to
stray light contamination from outside the field-of-view
(e.g. Gendreau 1995) and to a particle background
component which  increases with off-axis angle (Kubo et al. 1994).

 We use the Point Source Search (PSS) algorithm (Allan 1992)
to select candidate sources in the full 20 arcmin radius field-of-view,
down to a low level of significance
($3 \sigma$). PSS detects peaks above a given
threshold and fits the PSF to the observed surface brightness distribution
to decide whether these peaks are real sources or simply Poissonian
fluctuations. In addition, we run the PISA source detection algorithm
(Draper \& Eaton 1995)
to check whether any sources (especially confused or double sources)
have been missed by the PSS. Finally, we  include in our source list only
the sources, detected by either the above two algorithms, whose counts
in a detection cell of 1 arcmin  exceed the 4$\sigma$ background fluctuations.
At this level, only $\sim$0.1 spurious sources are expected in our survey.
At faint fluxes confusion may start posing problems.
A lower limit on the number of confused sources is found as follows.
The {\it Ginga} fluctuations \ns (Butcher et al. 1997) predicts a surface
density of
$\sim$50 $\rm deg^{-2}$ at the flux limit of our survey ($\sim 5\times10^{-14}$
\funits), translating to 0.014 sources per  1 arcmin radius beam or
$\sim 0.4$ double sources per field. Sources fainter than the flux limit
of our survey exacerbate the confusion. Of course, if the logN-logS
flattens from  Euclidean, as is the case in soft
X-rays, confusion problems will be relaxed.

A total of 26 point sources (there is no significant evidence for extension)
 were detected in our 3 \asca fields: 10 in the QSF3 field,
9 in the GSGP4 and 7 in the BJS855 field.
The flux limit in the QSF3 field is deeper, by about 30 per cent,
 compared to the other two fields. Hence, we expect to detect $\sim$50 per cent
more sources in QSF3, assuming an integral \ns slope of
$\gamma=1.5$;
 this translates to 10-14 sources in agreement with our observed number.
Therefore, there is no evidence for large field-to-field fluctuations
in the number of sources detected.
Note that the upper limit on the  fluctuations
in the 2-10 keV band from  the {\it HEAO-1} all-sky survey
is 5 per cent  on few degree scales (see Fabian \& Barcons 1992).

Count rates  were estimated as follows. In most cases,
 we measure the  source counts in
a 1 arcmin radius. This radius contains about 30 per cent of the source
counts on-axis. As most of our sources are faint, with less than 50 counts in
the 1 arcmin radius detection cell, use of a larger radius would
increase the source flux errors. For the few relatively bright sources, we use
a
radius of 2 arcmin.
 We then subtract the background counts as measured in a nearby
'source free' region. Count rates are calculated using the
exposure maps of the mosaic images;
The faintest source has a count rate of
$\sim 8\times 10^{-4}$ $\rm ct~s^{-1}$.

\subsection{The source list}

 We cross-correlate the \asca hard (2-10 keV) source positions with those
from the {\it ROSAT} PSPC (0.5-2 keV).
 These cross-correlations provide us immediately with the
optical identifications for most \asca sources, since a large fraction
($\sim75$ per cent) of our {\it ROSAT} survey sources have been
spectroscopically
identified.
18 \asca sources have counterparts in the $5\sigma$ {\it ROSAT} list
 (see Georgantopoulos et al. 1996), within 90 arcsec radius.
 As the rms error on the ASCA positions is $\sim 50$ arcsec (see below),
 we expect $\sim 95 $ per cent of our ASCA X-ray centroids to lie
 within  90 arcsec radius.
 Only three of the sources have two, $5\sigma$, {\it ROSAT} counterparts
 within the above radius. In these cases,  we assumed that the
 real counterpart is the nearest source; the details are given in table
3 below.
 Five more \asca sources
have counterparts in the deeper $4\sigma$ {\it ROSAT} list. Finally, three
hard X-ray sources have no {\it ROSAT} PSPC counterparts.
We note that due to the high density of {\it ROSAT} sources
 (typically $\sim$150 $\rm deg^{-2}$ at our faint flux limits) a few of the
 above cross-correlations may be chance coincidences, especially those
 at large separation. The cumulative number of {\it ASCA-ROSAT} (2-10 keV
 vs. 0.5-2 keV)
cross-correlations as a function of separation in arcsec is given in
table 2 for both the 5$\sigma$ and the 4$\sigma$ {\it ROSAT} lists.
The expected number of objects, assuming that the {\it ROSAT} sources are
distributed randomly with respect to the \asca sources, is given as well.
Note however, that the  above estimate of the number of random coincidences
 is  conservative  since we do not exclude the {\it ROSAT} sources that may
have a true \asca counterpart in the  calculation of the number density
of random ROSAT sources.
The above cross-correlation gives an rms error for the \asca GIS positions of
$\sim 50$ arcsec.

\begin{table}
\begin{center}
\caption{Cumulative number of {\it ASCA-ROSAT}  cross-correlations vs.
separation R}
\vskip 0.5cm
\begin{tabular}{ccccc}
\small
  & \multicolumn{2}{c}{ROSAT $5\sigma$} &
\multicolumn{2}{c}{ROSAT $4\sigma$} \\ \hline
$<R$ (arcsec) & Obs. & Exp. & Obs. & Exp.  \\ \hline
30       & 5 &0.7   &6 &        1.0\\
45       & 15&1.7&17& 2.2  \\
60       & 17&3.0&20& 3.9  \\
75       & 18& 4.5  &23& 6.1   \\
90       & 21&6.5&29& 8.5  \\
\end{tabular}
\end{center}
\end{table}

We give the list of sources detected in the hard 2-10 keV band in table 3.
The source name is given in
column (1); columns (2) and (3) give the \asca and { \it ROSAT} equatorial
(J2000)
coordinates for each object;  the offset between the \asca and {\it ROSAT}
positions is listed in column (4), in arcsec; column (5) contains the \asca
 GIS count
rate in the 2-10 keV band together with the photon errors in units of
($10^{-3}$ $\rm ct~s^{-1}$); columns (6) and (7) contain the
soft {\it ROSAT} PSPC and \asca GIS flux (1-2 keV) in units of $10^{-14}$
\funits.
 We converted the 1-2 keV count rates to  fluxes using a spectral
index of $\Gamma=1.7$ for all objects.
 Of course, this is not strictly true for all objects. However,
 the choice of spectral index
affects very little the resulting flux due to the very narrow
spectral band.
The conversion factors are then $2\times10^{-11}$ and $1.2\times 10^{-11}$
 $\rm erg~cm^{-2}~ct^{-1}$ for GIS and PSPC respectively.
If the object is not detected in the GIS 1-2 keV band down to the
$3\sigma$ detection threshold, the $3\sigma$ upper limit is quoted (see
Kraft, Burrows \& Nousek 1991).
 Finally column (8)
contains the optical identification and redshift where available.
 An outline of the optical observations and identification
procedure of our {\it ROSAT} survey are given in Georgantopoulos et al. (1996)
 while the full  details will be published elsewhere (Shanks et al, in
preparation).
We denote with asterisk (*) the sources detected in the low
significance ($4\sigma$) {\it ROSAT} list. The sources denoted with
question-mark
(?) were too faint optically (typically $B>22$) to give good signal-to-noise
optical spectra. Three sources were not spectroscopically observed
due to fibre positioning restrictions.
Note that there are appreciable differences between the {\it ROSAT} and
\asca
soft 1-2 keV flux. Although some variability in QSOs might be expected,
in other cases it could point towards  a possible
misidentification, as for example in the case of AXJ0057.6-2731.
 At low fluxes the  errors are expected to be
 significant:  typical errors from photon  statistics alone
are  of the order of 40 per cent for the faintest sources in the QSF3 field.

{}From table 3 we see that several  QSOs are detected; their mean
redshift is $z\approx1.1$. Two clusters are also detected. One of those,
AXJ0057.0-2741
is a high redshift cluster (z=0.561); its soft X-ray properties are
discussed in Roche et al (1995).
We have also identified a number of galaxies as potential
counterparts to the \asca sources.
One (AXJ1047.2-0028) is classified as an early type galaxy, on the basis of
absorption features in its optical spectrum; its redshift is
z=0.08 while its luminosity, $L_x\approx 2\times 10^{42}$ \lunits,
albeit high is not atypical of early-type galaxies detected by
Einstein and {\it ROSAT} (e.g. Fabbiano 1989).
 We have also identified  four  NELGs with luminosities ranging from
$L_x \sim 10^{42}$ to $10^{44}$ \lunits.
Despite the presence of several galaxies in our \asca survey, we note that  a
significant galaxy contribution to the hard X-ray  background cannot yet be
firmly established, due to the poor statistics. Most NELGs are faint
{\it ROSAT} sources ($<5\sigma$) and thus the possibility that some are due
to chance coincidences cannot be ruled out.

\subsection{The hardness ratios}

 Additional clues on the origin of the faint X-ray sources
come from their hardness ratios.
 Here, we define the {\it average}  hardness ratio as, $h-s/h+s$,
where $h$ and $s$ are the total number of counts in the detection cells,
in the 2-10 and 1-2 keV bands
respectively, for a given group of sources. A detailed analysis of the
combined \asca and {\it ROSAT} spectra is given in Georgantopoulos et al. (in
preparation).
The hardness ratio of all sources (excluding the star)
is 0.23$\pm0.04$. We convert the hardness ratios to photon indices
using XSPEC at a mean off-axis angle of 8 arcmin. The resulting
spectral index is $\Gamma=1.30\pm0.10$ ($1\sigma$ error).
The hardness ratio of the galaxies and unidentified  sources,
 i.e. excluding the identified QSOs, the two clusters and the star,
 has a value of 0.38$\pm$0.06 corresponding to an  index of
$\Gamma=0.92\pm0.16$.
This is flatter than  the spectral index of the 2-10 keV XRB,
 which has $\Gamma\sim 1.4-1.5$ (Gendreau et al. 1995, Chen, Fabian
\& Gendreau 1997).  Hence, these objects may be the first faint
examples of the hard spectrum population that makes a
substantial contribution to the hard XRB.
In contrast, the
average QSO hardness ratio is $0.04\pm$0.06 yielding a spectral index
of $\Gamma=1.78\pm0.16$, marginally flatter than  the average {\it ROSAT} QSO
spectral index in
 our fields (Stewart et al. 1994) but similar to the average
 nearby AGN spectrum in this band (Nandra \& Pounds 1994).
Chen et al. (1997) present \asca SIS observations of the two bright
QSOs in the QSF3 field.
The combined SIS+PSPC fits give spectral indices of $\Gamma\approx 3.1
\pm 0.1$.
 This spectral index is considerably steeper than ours,
 possibly due to the lower energy
 range  of the \asca SIS and {\it ROSAT} PSPC which can be affected by soft
excesses
 in the QSO spectra.
 However, both their work and ours suggest  that the average QSO spectra
 are steeper than that of the XRB.
 This result, the spectral paradox, was noted earlier with {\it HEAO-1}
 (e.g. Boldt 1987) and \ginga albeit at much brighter fluxes ($>7\times
 10^{-12}$ \funits).

\section{THE NUMBER-COUNT DISTRIBUTION}

\subsection{The 2-10 keV logN-logS}

 We calculate the extragalactic number-count distribution,
\ns, in the 2-10 keV band. We use  the 25 sources detected in
our three fields, excluding only the  star in the QSF3 field.
Due to the strong vignetting of the  XRT, the faintest sources can only
be detected in the center of the GIS field-of-view, where the
sensitivity is the highest, while the bright sources can be
detected at all off-axis angles. Therefore, we need first to
estimate the sky coverage of our survey. The cumulative area covered
 as a function of the limiting flux is given in Fig. 1.


\begin{figure}
\plotfiddle{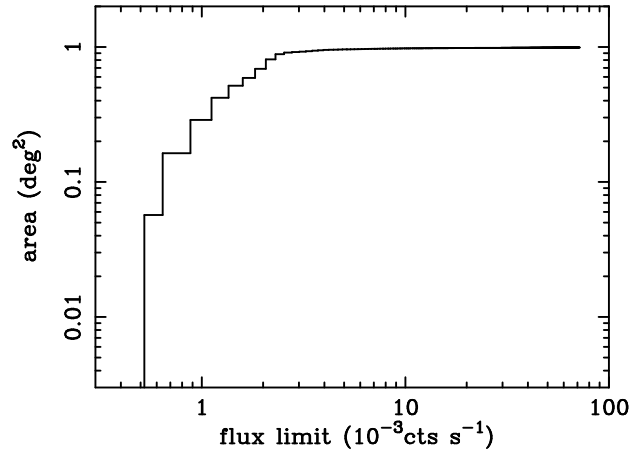}{200pt}{-90}{40}{40}{-150}{240}
\caption{
The sky coverage of our survey as a function of limiting count rate.}
\label{fig:fig1}
\end{figure}


The integral number-counts, $N(>S)$, are given by the sum
$\Sigma (1/\Omega_i)$, where $\Omega_i$ is the area coverage
at the flux, $S_i$, of the source $i$.
To facilitate comparison with previous results we use
a spectral index of $\Gamma=1.7$; this corresponds to a count-rate-to-flux
conversion factor of $5.8\times10^{-11}$ $\rm erg~cm^{-2}~ct^{-1}$;
 we note that the count-rate-to-flux conversion factor for our mean  spectral
index of $\Gamma\sim 1.3$ would be  $\approx 6.5$ $\rm erg~cm^{-2}~ct^{-1}$.
 The resulting \ns is plotted in Fig. 2 (histogram).

\begin{figure}
\plotfiddle{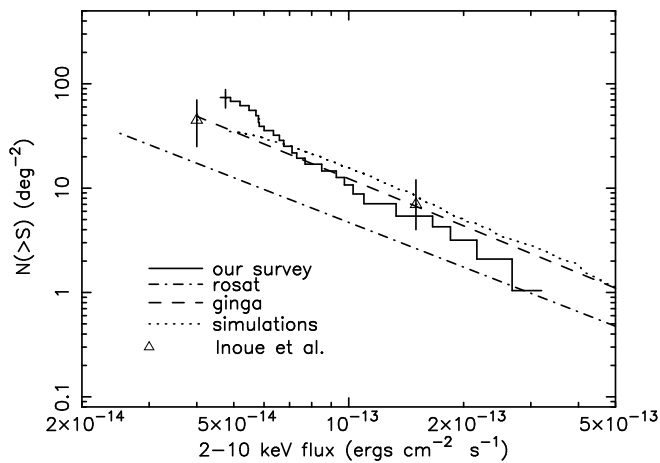}{200pt}{-90}{35}{35}{-130}{210}
\caption{The derived integral \ns in the 2-10 keV band from our
survey (histogram).
Also shown the {\it ROSAT} counts (dot-dashed line), converted in the
2-10 keV band using $\Gamma=2$, the \ginga counts (dashed
line), the \ns derived from the 100 simulated ASCA fields (dotted line).
The triangles are adapted from the Inoue et al. deep ASCA survey.
All errors correspond to the $1\sigma$ confidence level.}
\label{fig:fig2}
\end{figure}
The preliminary number-counts from two deep, Japanese, \asca surveys
 are adapted from Inoue et al. (1996)  (triangles).
We also plot the soft (0.5-2 keV) number-counts (dot-dash line), as derived
from
our {\it ROSAT} survey (Georgantopoulos et al. 1996), converted to the
2-10 keV band using  a power-law index of $\Gamma=2$ for the {\it ROSAT}
source spectra (Hasinger et al. 1993, Vikhlinin et al. 1995b).
The \ns measured from the \ginga fluctuations (Butcher et al. 1997),
 is plotted as a dashed line.
Finally, the dotted line gives the \ns derived from 100 Monte Carlo
simulations of  \asca fields (see below). All errors plotted corerspond to
the $1\sigma$ significance level.
Inspection of Fig. 2 suggests the following.
The \ns of our \asca survey appears to be in
 rough agreement with the Japanese  \asca surveys, especially at bright fluxes.
Furthermore, the \asca \ns is in
agreement with the number-counts measured from the
\ginga fluctuations. Instead, the \asca number-counts lie  significantly above
the
{\it ROSAT} \ns. This excess number density of hard X-ray  sources
suggests that a new source population, other than the QSOs which dominate
the soft \ns, is present in our \asca survey.
This population could remain
undetected in the {\it ROSAT} surveys of  comparable flux depth
($S_{\rm 0.5-2 keV}>10^{-14}$ \funits)  due to its flat or absorbed X-ray
spectrum.
However, we have first to rule out any  possibility that systematic effects
could alter the \ns form and produce the observed excess density.
Such effects in the
source detection and flux estimation are examined in the next section.

\subsection{Checking for systematic effects}

The \ns derived above may be affected by several systematic effects in
the source detection and flux calculation procedure.
The most important are the Eddington bias and source confusion.
The Eddington bias is the net gain of sources near the
flux limit of the survey due to flux errors.
Murdoch, Crawford \& Jauncey (1973) and Schmitt \& Maccacaro
(1986) discuss this effect and give analytic corrections
for pure power law counts. However, the above corrections assume that
the flux error distribution is well determined.
The Eddington bias is going to have a small  effect
in our \ns estimation, either if the flux errors are
negligible or alternatively, if the \ns breaks to a flatter
than Euclidean power law, as in the case of the {\it ROSAT} number-counts.
Source confusion plays an important role at faint  fluxes and
may result in either the  increase or the decrease of the total number of
sources
detected.
If the confused  sources  are below the detection threshold,
then the  merged source may appear above the survey's
flux limit and thus we end up with a net gain in the
number of sources. Alternatively, two sources above the
detection threshold could merge to form a brighter source,
thus resulting in a loss of fainter sources.

We  check the validity of our \ns using two tests.
 We first derive the soft (1-2 keV), \ns, from  our three \asca
GIS fields. Comparison with the well-determined {\it ROSAT}
\ns then provides powerful constraints on possible
GIS systematic flux errors.
Using the detection methods described earlier in this paper,
 we detect 15 sources in our three fields (of which two are identified as
stars), in the 1-2 keV band
down to a flux limit of $\sim 10^{-14}$ \funits.
The integral \ns for the 13 sources, excluding the two stars, is plotted in
Fig. 3.
It is compared with the extragalactic 0.5-2 keV
{\it ROSAT} \ns (Georgantopoulos et al. 1996) converted to the 1-2 keV
band using a spectral index of
$\Gamma=2$. Despite the poor statistics of the
\asca counts,  we see that the two \ns are in good agreement,
demonstrating that the combined effects of flux errors and source confusion do
 not significantly change  the \ns at soft fluxes.


\begin{figure}
\plotfiddle{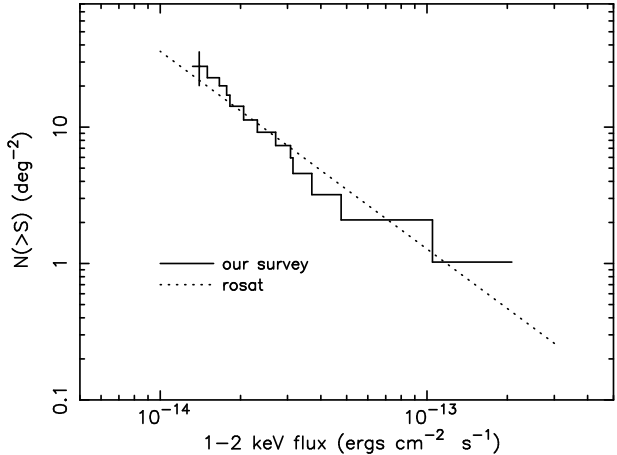}{200pt}{-90}{35}{35}{-130}{210}
\caption{The integral \ns of the ASCA soft sources detected in the 1-2 keV
band (histogram) compared to the {\it ROSAT} 1-2 keV \ns.}
\label{fig:fig3}
\end{figure}


As an additional test, we performed Monte Carlo simulations of the 2-10 keV
images. We create 100 fields in total, having the same exposures times,
and background count rates as the three observed fields. In each
field X-ray sources were assigned random positions, while
their input fluxes were drawn from an integral \ns with Euclidean
slope ($\gamma =1.5$) and a normalization of
$ 390 ~\rm deg^{-2}$ at $10^{-14}$ \funits in agreement with  the  \ginga
counts.
The faint limit of our simulation is
$5\times 10^{-15}$ \funits, about an order of magnitude below the
flux limit of our \asca survey; at this flux level the \ns saturates the
2-10 keV XRB. A uniform particle component is added
with a count rate of $5\times 10^{-5}$ $\rm ct~s^{-1}$ (e.g. Kubo et al.
1995). For each source the photons are spread using the actually
measured PSF. Finally, the vignetting of the XRT (Serlemitsos et al. 1995)
has been applied analytically. We apply  exactly the same
detection procedure, used in the case of the actual fields.
We detect $\sim815$ sources in total. The integral \ns of the simulated fields
was given in  Fig. 2 (dotted line).
We see that there is rough  agreement between the simulations and the input
\ns.  This does not imply that flux errors are  negligible.
 Our  simulations show that flux errors as large as
100 per cent  can be expected  near the survey's flux limit.
It is therefore probable that the observed agreement between
the simulated and input \ns near the survey's flux limit
is because confusion approximately cancels out the Eddington
bias effect.

\subsection{Contribution to the XRB}

The summed flux of the 10 sources in the deepest field, QSF3,
is $\sim 8\times10^{-13}$ \funits in the 2-10 keV band.
The total XRB flux in the same band is $\sim 6.5\times 10^{-12}$
$\rm erg~cm^{-2}s^{-1}~deg^{-2}$ (Chen et al. 1997).
This translates to a resolved source contribution of 12 per cent.
However, this is only a lower limit since it does not take into
account the strong telescope vignetting which prevents the
detection of faint sources at large off-axis angles.
 Instead, we need to estimate  the source contribution by using the
observed \ns distribution.
We fit a power law model ($dN \propto N^{-\beta}$) to the differential
source counts (e.g. Murdoch et al. 1973). We find a slope of
$\beta=3.27\pm0.57$ (integral slope of 2.27),
where the errors quoted correspond to the 90 per cent confidence level.
We note that a Euclidean slope cannot be ruled out  at the $\sim2\sigma$
confidence level. Given the limited statistics and the small
flux range covered by our survey, we conclude that there is  no
strong evidence yet for a non-Euclidean count distribution.
Fixing the integral \ns slope to the Euclidean value of 1.5 gives
a normalization of
4.4$\times10^{-19}$ $\rm deg^{-2}~(erg~cm^{-2}~s^{-1})^{1.5}$
 comparable to  the \ginga normalization. Using  the above
\ns ($\gamma=1.5$), we estimate a source contribution of $\sim13$
 $\rm keV~cm^{-2}~s^{-1}~sr^{-1}$
 (2-10 keV) down to the limiting flux of our survey or about 30 per cent
of the observed XRB in this band. Extrapolation of our counts  down to
$5\times 10^{-15}$ \funits,
i.e. an order of magnitude fainter than the limiting flux of the
present survey and at comparable flux depth to the deep {\it ROSAT} surveys,
 produces all  the XRB.
Our best fit slope of $\gamma=2.27$ gives a contribution of
over 40 per cent down to the flux limit of our survey
while it saturates the XRB at a flux of $\sim 2\times 10^{-14}$
 \funits.
The above calculations show  that future X-ray missions like
{\it JET-X, XMM} and {\it AXAF} will be able to resolve  the hard XRB,
unless the counts turn over to a flatter
slope, at fluxes fainter than the flux limit of our survey.

We finally note that although {\it HEAO-1} and \asca observations have shown
that
the slope of the hard XRB is in the range $\Gamma=1.4-1.5$ in the 1-10
keV band, its exact normalization  is not yet well determined.
In our calculations above, we use the measurements
of the XRB from the \asca SIS observations of the QSF3 field (Chen et
al. 1997). These give a normalization of $10.5\pm0.4$ \xrbunits, at 1
keV, consistent with our  {\it ROSAT} XRB measurements of the same
field (Georgantopoulos et al. 1996) and other {\it ROSAT} fields.
 \asca GIS observations  of various fields (Ishisaki 1996)
give similar values for the normalization.
 However, \asca SIS observations (Gendreau et al. 1995)
 yield a somewhat lower value for the
XRB (9 \xrbunits at 1 keV) closer to the {\it HEAO-1} measurements
(Marshall et al. 1980).
If we use instead the Gendreau et al. (1995) value, our
number-count distribution ($\gamma=1.5$) saturates
the XRB at even higher  fluxes  ($7\times 10^{-14}$ \funits).

\section{CONCLUSIONS}

We have discussed deep \asca GIS observations of three fields from the
deep {\it ROSAT} survey of Shanks et al. (in preparation).
We detected 26 sources down to a limiting flux of
$\sim 5 \times 10^{-15}$ \funits (2-10 keV). In the deepest field these
sources contribute about 30 per cent of the XRB, as measured by Chen et
al. (1997).
 There appears to be an excess density of
hard X-ray sources of about a factor of three above the {\it ROSAT} counts.
The agreement of the \asca and {\it ROSAT} soft (1-2 keV) \ns as well as
Monte Carlo simulations suggest that this excess is not an artefact of
flux errors or  confusion at faint fluxes. Our finding confirms
and extends previous {\it HEAO-1} and \ginga results at much brighter fluxes.
The observed hard X-ray source excess density suggests that a population,
other than the QSOs that dominate the soft (0.5-2 keV) source counts,
remains unidentified at hard X-rays. This population must have
a flat or absorbed X-ray spectrum since it is not detected in
the {\it ROSAT} band  at comparable, bright, flux levels i.e.  $>10^{-14}$
\funits.
Indeed,  previous spectroscopic  observations of our ROSAT survey
at the Anglo-Australian Telescope show a relatively low fraction of QSOs
among the hard X-ray sources:  we detect  8  QSOs,
2  clusters, 1  star
while 6 sources coincide with NELGs or early-type galaxies. The remaining
 9 sources are unidentified.
 Although we cannot yet conclusively rule out the possibility that some  of the
 9 unidentified sources are broad-line AGN, the average hardness
ratio of the 15  galaxies and unidentified sources
 yields a spectral index of $\Gamma=0.92 \pm 0.16$,
significantly different from the QSO spectral index ($\Gamma\approx
1.78\pm0.16$) and flatter
 than  the XRB spectral index ($\Gamma\sim 1.4$).
This corroborates the presence of a new flat spectrum population that
could produce a large fraction of the hard XRB.
Nevertheless, the nature of this population remains unknown.
{\it ROSAT} observations have detected a large number of NELGs at faint
fluxes (Roche et al. 1995, Griffiths et al. 1995, 1996, Boyle et al.
1995), which may be  associated with obscured active
nuclei or starforming galaxies.
Narrow-line type 2 AGN are also detected in \asca surveys (e.g. Ohta et
al. 1996).
In our \asca survey we  find  a relatively
large number of NELGs and early-type galaxies.
However, we emphasize yet again that due to the large positional error
box of the \asca detectors, a few of these identifications may due to chance
coincidences. Therefore, the amount of the NELG contribution at hard
X-rays  remains yet unknown.

In conclusion, our \asca survey has succeeded in resolving and identifying
a large fraction ($\sim 30 $ per cent) of the hard 2-10 keV XRB.
Although some QSOs are  detected, \asca has clearly detected
 another population with a flat hard X-ray spectrum.
  Although there are hints that this could be associated with  NELGs,
  the limited statistics of the present survey, together with the large
positional
errors of the \asca GIS hinder the identification of this new population.
Further \asca or {\it SAX} observations, of fields previously observed by {\it
ROSAT},
together with spectroscopic follow-up observations of the unidentified
sources are necessary to  clarify the nature of the hard X-ray population.

\vfil
\eject

{\bf \large ACKNOWLEDGEMENTS}

  We are grateful to Ian Hutchinson for his help on the
 \asca simulation software. We would like to thank the
 referee Xavier Barcons for many useful comments and suggestions.
 This research has made use of data obtained through the
 LEDAS and HEASARC online services, provided by the
 Leicester University and the NASA Goddard Space flight Center
 respectively.
 The optical data were obtained at the
 Anglo-Australian telescope. IG, GCS and OA acknowledge the
 support of PPARC.

\end{document}